\documentclass[aps,pra,twocolumn,amsmath,amssymb,floatfix,showkeys,showpacs]{revtex4-2}
\usepackage{bm} % bold math
\usepackage{color}
\usepackage{graphicx}
\usepackage{epsfig}
\usepackage{amssymb}
\usepackage{amsmath} %,amssymb}
\usepackage{amsthm}
\usepackage{amstext}
\usepackage{dcolumn} % Align table columns on decimal point
\usepackage{makeidx}
\usepackage{subfigure}
\usepackage{xcolor}
\definecolor{darkgreen}{rgb}{0,.5,0}%testing coloring
\usepackage[colorlinks,urlcolor=blue,filecolor=blue,linkcolor=blue,citecolor=blue]{hyperref}

\begin{document}

\title{Optomechanical systems with nonlinear interactions: photon blockade, collapse-revival effect and Fano-like resonance}
\author{A.P. Saiko$^{1}$}\email{saiko@physics.by}
\author{G.A. Rusetsky$^{1}$}
\author{S.A. Markevich$^{1}$}
\author{R. Fedaruk$^{2}$}
\affiliation{$^{1}$Scientific-Practical Material Research Centre, Belarus National Academy of Sciences, 19 P.Brovka str., Minsk 220072 Belarus}
\affiliation{$^{2}$Institute of Molecular Physics, Polish Academy of Sciences, Smoluchowski Str. 17, 60-179 Poznan, Poland}
\date{\today}

\begin{abstract}
Closed-form expressions for the average amplitude of the optical field in optomechanical systems are obtained, in which, in addition to the linear interaction, quadratic and cubic interactions of the vibrational mode of the mechanical resonator with the mode of the optical resonator are considered. In the framework of the non-secular perturbation theory, using the Bogoliubov averaging method, it is shown that the effects of photon blockade, collapse and revival of optical oscillations in such systems can be realized. The main contribution to the formation of revivals is provided by the Kerr self-action of the optical mode and the cross-Kerr interaction of the fourth degree in optical and mechanical amplitudes. The cross-Kerr interactions of the sixth- and eighth-order in amplitudes destroy the regular structure of revivals. The influence of these cross-Kerr nonlinearities disappears with an increase in the decay rate of the optical mode and is also completely suppressed at zero temperature. It is shown that the asymmetry of the spectral line of the optical field intensity in the cavity is most pronounced with an increase in the degree of nonlinearity and is explained by Fano interference.
\end{abstract}

%\keywords{nonlinear optomechanical system, photon blockade, collapse-revival effect, Fano-like resonance}

\maketitle
\section{Introduction}
In optomechanical systems, light pressure causes mechanical resonator (oscillator) vibrations, which in turn control the behavior of the intra-cavity optical mode \cite{pp1}. Optomechanical interactions can result in cooling \cite{pp2, pp3, pp4} and amplification of vibrational modes of the mechanical resonator \cite{pp5}. At the strong driving on the cavity mode, the multi-photon strong coupling regime is realized and many observed physical phenomena, including optomechanically induced transparency \cite{pp6, pp7, pp8, pp8a}, normal-mode splitting \cite{pp9, pp10}, quantum entanglement \cite{pp11, pp12, pp13, pp13a}, photon blockade effect \cite{pp14, pp14a} and quantum state transfer \cite{pp15, pp16, pp17}, can be understood using a linear description \cite{pp1}. Macroscopic mechanical entanglement in two distant optomechanical systems has been investigated in \cite{pp18}. A number of papers have been devoted to taking into account nonlinearities (cubic or fourth degree) of mechanical oscillators in optomechanical systems, including the study of the anharmonicity of a quantum oscillator \cite{pp19, pp20}, as well as the study of such effects as the generation of second-order sidebands \cite{pp21}, steady-state mechanical squeezing \cite{pp22}, normal mode splitting \cite{pp23}, optomechanically induced transparency \cite{pp24, pp25}, optomechanical entanglement \cite{pp26} and formation of Kerr and cross-Kerr nonlinearities \cite{pp27, pp27a}.
Optomechanical systems with a coupling proportional to the squared mechanical displacement have been studied, in which collapses and revivals of mechanical motion have been described \cite{pp28, ppAmit}.  It has recently been shown that an optomechanical system without the quadratic nonlinear interaction can also display the collapse-revival effect \cite{ppDozal, ppUrzua}.

The combination of micron-scale or smaller mechanical resonators with optical resonators and superconducting qubits is currently one of the foundations for the development of quantum devices. These devices are unique tools for fundamental experiments at the cutting edge of quantum mechanics. The operation of such optomechanical systems is mainly based on the standard linear optomechanical Hamiltonian, where the coupling between the mechanical and optical resonators is linear in mechanical displacements  $x$. However, the optomechanical interaction is nonlinear in nature from the beginning, and to solve some fundamental problems it is necessary to take into account quadratic ($ x^{2} $) and even cubic ($x^{3}$) \cite{pp29} mechanical displacement interactions. Microwave optomechanical experiments performed in the self-oscillation regime have demonstrated that the limit cycle dynamics of such a system is sensitive to nonlinearities in the optomechanical coupling \cite{pp29}. Explaining the experimental results required going beyond the standard model of an optomechanical system. It was shown that these results can be quantitatively explained only by adding quadratic and cubic interactions to the standard linear optomechanical Hamiltonian. In this case, the contribution of thermo-optical nonlinearities was excluded. The theoretical analysis in \cite{pp29} was based on the use of Heisenberg equations in the multi-quantum approximation, i.e. in these equations a transition was made from the operators of optical and vibrational modes to their classical analogues and noise sources were neglected. Until now, purely quantum aspects of the behavior of such a nonlinear optomechanical system, when the external exciting photon field is weak, remain unexplored. Thereby, the following issues are topical: (i) finding the energy spectrum of the system, (ii) studying the features of the nonlinear dynamics of optical and vibrational modes, which can manifest themselves in the collapse-recovery effect of their oscillations, and (iii) establishing a direct connection between the dissipative dynamics of the optomechanical system and the physics of Fano resonances. In our study, the listed problems will be solved (under the conditions specified below) by constructing an approximately diagonal effective Hamiltonian within the framework of non-secular perturbation theory using Bogoliubov averaging in its canonical formulation. We demonstrate how successive addition of nonlinear (quadratic and cubic) interactions modifies the collapse-revival structure. This structure changes dramatically: the time of revival appearance is reduced, and each revival is blurred and acquires well-defined satellites. We show that the formation of revivals is provided by the Kerr self-action of the optical mode and the cross-Kerr interaction of the fourth degree in optical and mechanical amplitudes. At the same time, the cross-Kerr interactions of the sixth- and eighth-order in amplitudes destroy the regular structure of revivals. Moreover, the influence of these cross-Kerr nonlinearities disappears with an increase in the decay rate of the optical mode and is also completely suppressed at zero temperature. It is shown that the asymmetry of the spectral line of the optical field intensity in the cavity is most pronounced with an increase in the degree of nonlinearity and is explained by Fano interference. The interest in studying such optomechanical systems is due to the fact that they can be used as extremely sensitive detectors of mass, force and position \cite{pp1}. Strong optomechanical interactions at the level of individual photons can be used in nano-optomechanical devices to induce controlled nonlinear couplings between individual photons and phonons, with potential applications in quantum information processing \cite{pp30}.

The remainder of this paper is organized as follows. In Sec. II we introduce the model of  optomechanical systems with nonlinearities in optomechanical interactions up to the third order in the amplitude of mechanical oscillations, derive its effective Hamiltonian and corresponding energy spectrum. Peculiarities of dissipative dynamics of the model under consideration are described and discussed in Sec. III.  In Sec. IV it is shown that the spectral line of the optical field intensity in the cavity has a pronounced asymmetric shape, which is a characteristic feature of Fano interference and indicates that the dissipative dynamics of the optomechanical system can be related to the physics of Fano resonances. Finally, we conclude with a brief summary in Sec. V.

\section{Effective Hamiltonian and energy spectrum of the optomechanical system}

Let us consider an optomechanical system in which, in addition to linear and quadratic interactions, the cubic interaction manifests itself quite significantly. Physical processes in such a system can be described by the following Hamiltonian \cite{pp29}:\textbf{}

\begin{equation} \label{GrindEQ1}
H=H_{0}+V+V_{drive}^{c} ,
\end{equation}
\[
H_{0} = \omega_{c} a^{\dag} a + \Omega b^{\dag} b,
\]
\[V=-a^{\dag } a\sum _{n=1,2,3}g_{n} (b^{\dag } +b)^{n},\]
\[
V_{drive}^{c} = i\varepsilon (a^{\dag} e^{-i\omega_{d} t} - H.c.),
\]

\noindent where  $ \omega _{c}  $  is the frequency of the optical resonator,  $ \Omega  $  is the frequency of mechanical oscillations,  $ g_{n} $  is the optomechanical coupling strength, \textit{n}= 1,\,2 and 3 for linear, quadratic and cubic interactions, respectively;  $ a $  and  $ b $  represent the respective annihilation operators of photon and vibrational quanta (Planck's constant  $ \hbar  $  is taken equal to 1). The term  $ V_{drive}  $  describes the interaction with the external exciting photon field of the amplitude $ \varepsilon  $  and the frequency  $ \omega _{d}  $. The transition to a rotating coordinate system for the field in the optical resonator using the evolution operator  $ U^{c} =e^{i\omega _{d} a^{\dag } at}  $  leads to the time-independent excitation term  $ i\varepsilon (a^{\dag } -a) $  and the appearance of the resonance detuning  $ \Delta =\omega _{c} -\omega _{d}  $.

Let's consider the case where  $ \varepsilon =0 $.  Then, in the interaction representation, the Hamiltonian \eqref{GrindEQ1} is written as:
\[V(t)=e^{i(\Delta a^{\dag } a+\Omega b^{\dag } b)t} Ve^{-i(\Delta a^{\dag } a+\Omega b^{\dag } b)t} =\]
\begin{equation} \label{GrindEQ2}
-a^{\dag } a\sum _{n=1,2,3}g_{n} (b^{\dag } e^{i\Omega t} +be^{-i\Omega t} )^{n}  .
\end{equation}
Since for real optomechanical systems the inequalities $\Omega {\rm \gg }g_{1} ,g_{2} ,g_{3} $ are fulfilled, we can use the method of the non-secular perturbation theory for averaging over fast oscillations $e^{\pm in\Omega t} $ (where $n=1,2,3$). In the canonical form it can be realized using the Bogoliubov averaging method \cite{pp31, pp32, pp34}. In result, we obtain  up to the second order in small parameters $g_{1} /\Omega $, $g_{2} /\Omega $, and $g_{3} /\Omega $ the following effective diagonal Hamiltonian $H_{eff} $:
\begin{equation} \label{GrindEQ3}
H\to H_{eff} =\Delta a^{\dag } a+\Omega b^{\dag } b+V_{eff}^{(1)} +V_{eff}^{(2)} ,
\end{equation}
where
\[V_{eff}^{(1)} =<V(t)>,\]
\begin{equation} \label{GrindEQ4}
V_{eff}^{(2)} =\frac{i}{2} \langle [\int _{}^{t}d\tau (V(\tau )-<V(\tau )>),V(t) ]\rangle ,
\end{equation}
Here the symbol $\left\langle ...\right\rangle $ means the operation of time averaging $\langle O(t)\rangle =\frac{\Omega }{2\pi } \int _{0}^{{2\pi \mathord{\left/ {\vphantom {2\pi  \Omega }} \right. \kern-\nulldelimiterspace} \Omega } }O(t)dt $ over fast oscillations $e^{\pm in\Omega t} $, square brackets denote the commutation operation, and \textit{t} in  the upper limit of the indefinite integral is the variable on which the result of integration depends. Substituting Eq. \eqref{GrindEQ2} into Eq. \eqref{GrindEQ4} and performing all necessary operations, we obtain:

\begin{equation} \label{GrindEQ5}
V_{eff}^{(1)} =-\frac{1}{2} g_{2} a^{\dag } a-g_{2} a^{\dag } ab^{\dag } b,
\end{equation}
\[V_{eff}^{(2)} =-\Lambda _{1} a^{\dag } aa^{\dag } a-\Lambda _{2} a^{\dag } aa^{\dag } ab^{\dag } b-\Lambda _{3} a^{\dag } aa^{\dag } ab^{\dag } bb^{\dag } b,\]

\noindent where
\[\Lambda _{1} =\frac{1}{\Omega } \left(g_{1}^{2} +g_{1} g_{3} +\frac{1}{4} g_{2}^{2} +\frac{1}{18} g_{3}^{2} \right),\]
\begin{equation} \label{GrindEQ6}
\Lambda _{2} =\frac{1}{\Omega } \left(2g_{1} g_{3} +\frac{1}{2} g_{2}^{2} +\frac{10}{12} g_{3}^{2} \right),\, \, \, \, \, \, \Lambda _{3} =\frac{10}{12\Omega } g_{3}^{2} ,
\end{equation}
$\Lambda _{1}$  is the Kerr parameter,  $g_{2}$,  $ \Lambda _{2}$  and  $\Lambda _{3}$  are the cross-Kerr parameters of the fourth, sixth and eighth degrees, respectively, in the amplitudes of the optical field and mechanical vibrations. It is evident from Eq. \eqref{GrindEQ5} that in the first order of non-secular perturbation theory, the quadratic interaction ($g_{2} $) contributes to the renormalization of the optical field frequency and is responsible for the appearance of the diagonal cross-Kerr interaction of the fourth degree in the optical and mechanical amplitudes. In the Kerr self-interaction of the optical field with the parameter $\Lambda _{1} $, the contributions of all three optomechanical interactions, proportional to $g_{1}^{2} $, $g_{2}^{2} $, and $g_{3}^{2} $, are presented separately, and there is also the cross term $\sim g_{1} g_{3} $. The parameter $\Lambda _{2} $ of the cross-Kerr interaction, the sixth-order in oscillation amplitudes, depends on $g_{2}^{2} $, $g_{3}^{2} $ and $g_{1} g_{3} $, while the highest order cross-Kerr parameter $\Lambda _{3} $ is determined solely by the cubic optomechanical interaction with the coupling strength $g_{3} $.

The eigenvalues of the Hamiltonian  $ H_{eff}  $  for the Fock states  $ {\left| n,m \right\rangle} ={\left| n \right\rangle} {\left| m \right\rangle}  $, where  $ n(m) $  denotes the number of photons (phonons), are  $ H_{eff} {\left| n,m \right\rangle} =E_{n,m} {\left| n,m \right\rangle} , $

\begin{equation}
 \label{GrindEQ7}
  \begin{split}
  E_{n,m} = & (\Delta -\frac{1}{2} g_{2} )n+\Omega m-g_{2} nm  \\ & -\Lambda _{1} n^{2} -\Lambda _{2} n^{2} m-\Lambda _{3} n^{2} m^{2}.
  \end{split}
\end{equation}

There is a strong nonlinearity of eigenvalues  $ E_{n,m}  $  of the Hamiltonian  $ H_{eff}  $  from photon and phonon filling numbers and their products. This nonlinearity in the energy spectrum is the physical reason for the realization of the photon blockade effect in this optomechanical system.

\begin{figure}[ht] \centering \includegraphics[width=8 cm]{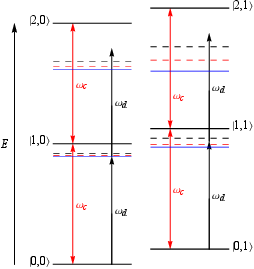} \caption{Spectrum  $ E_{n,m}  $  of the effective Hamiltonian  $ H_{eff}  $  obtained from Eq. \eqref{GrindEQ7}. Black solid levels show the spectrum without taking into account the interactions in the optomechanical system. Black dashed, red dashed and blue lines depict the spectrum taking into account the nonlinearity of the first order, the first and second orders, and the first, second and third orders, respectively. Arrows show the frequencies  $ \omega _{d}  $  and  $ \omega _{c}  $  of the exciting photon field and the optical resonator, respectively.} \label{fig:Fig1} \end{figure}

As an example, Fig. \ref{fig:Fig1} shows the spectrum \eqref{GrindEQ7} of the Hamiltonian  $ H_{eff}  $  of our model system without and taking into account the nonlinearities in the optomechanical system.
If driving field is on resonance with the  $ {\left| 0,0 \right\rangle} \to {\left| 1,0 \right\rangle}  $  transition (or with the transition  $ {\left| 0,1 \right\rangle} \to {\left| 1,1 \right\rangle}  $, when one vibrational quantum also participates in quantum transitions), the transition  $ {\left| 1,0 \right\rangle} \to {\left| 2,0 \right\rangle}  $  (or $ {\left| 1,1 \right\rangle} \to {\left| 2,1 \right\rangle}  $) should occur with a significant detuning  $ g_{2} /2+3\Lambda _{1}  $  (or $ 3g_{2} /2+3(\Lambda {}_{1} +\Lambda {}_{2} +\Lambda {}_{3} ) $) and therefore will be suppressed if the detuning significantly exceeds the value of the damping rate  $ \kappa  $  of the optical mode. Therefore, a photon blockade is realized for the absorption of the second photon. We here limit ourselves to this simple picture of the photon blockade, based on the consideration of the energy structure of the optomechanical system, without resorting to the standard procedure for calculating the two-photon equal-time correlation function in such cases.

\section{Dissipative dynamics of the system: collapses and revivals}

In this section, the evolution of the optomechanical system excited at the initial moment of time by a short pulse of an optical field with a carrier frequency $\omega _{d} $ is considered. That is, we will analyze the scenario when at the initial time the optical mode is populated in a monochromatic coherent state  $ {\left| \alpha  \right\rangle}  $, and the mechanical oscillator (resonator) is in a thermal state at temperature  $ T $, i.e., the wave function of the system at time  $ t=0 $  can be represented as  $ {\left| \psi (0) \right\rangle} ={\left| \alpha  \right\rangle} \sum _{m=0}^{\infty }p_{m}  {\left| m \right\rangle}  $, where  $ {\left| \alpha  \right\rangle} =e^{-\frac{{}^{\left|\alpha \right|^{2} } }{2} } \sum _{n=0}^{\infty }\frac{\alpha ^{n} }{\sqrt{n!} }  {\left| n \right\rangle}  $,  $\left|p_{m} \right|^{2} ={\bar{m}^{m} }/{(\bar{m}+1)^{m+1} }  $, and  $ \bar{m}=\left[\exp (\Omega /k_{B} T)-1\right]^{-1}  $ \textbf{ }is the\textbf{ }average thermal population of the mechanical oscillator at temperature  $ T $. To describe the dissipation of the optical mode and mechanical oscillations in the analytical calculation of the evolution of the optomechanical system, we will use the non-Hermitian effective Hamiltonian

\begin{equation} \label{GrindEQ__8_}
H_{eff}^{*} =H{}_{eff} -i\frac{\kappa }{2} a^{\dag } a-i\frac{\gamma }{2} b^{\dag } b,
\end{equation}

\noindent where $ \kappa  $  and  $ \gamma  $  are the damping rates of the optical mode and mechanical oscillations, respectively. Then the wave function of the optomechanical system at some time  $ t $  can be expressed as  $ {\left| \psi (t) \right\rangle} =\exp (-iH_{eff}^{*} t){\left| \psi (0) \right\rangle}  $  and the average amplitude of the optical field in the resonator can be calculated  $ \left\langle a\right\rangle ={\left\langle \tilde{\psi }(t) \right|} a{\left| \tilde{\psi }(t) \right\rangle}  $, where  $ {\left| \tilde{\psi }(t) \right\rangle} =N(t)^{-1/2} {\left| \psi (t) \right\rangle}  $, and  $ N(t)=e^{-\left|\alpha \right|^{2} } \exp (\left|\alpha \right|^{2} e^{-\kappa t} )\left[1+\bar{m}(1-e^{-\gamma t} )\right]^{-1}  $  is the normalization factor. Finally, we obtain

\begin{multline} \label{GrindEQ__9_}
%\begin{aligned}
\left\langle a \right\rangle = \frac{\alpha }{N(t)} \sum_{n,m=0}^{\infty} \left| p_{m} \right|^{2} e^{-|\alpha|^{2}} \frac{|\alpha|^{2n}}{n!} e^{-i\Delta t}  \\  \times \exp \left\{ i \left[ g_{2} \left( m + \frac{1}{2} \right)\right.\right.\\
+(\Lambda_{1} + \Lambda_{2} m + \Lambda_{3} m^{2})(2n + 1) \bigg] t \bigg\} e^{- \left[ \kappa \left( n + \frac{1}{2} \right) + \gamma m \right] t}.
%\end{aligned}
\end{multline}

\noindent Eq. \eqref{GrindEQ__9_} can be transformed by summing over  $ n $ :
\begin{equation}
\label{GrindEQ__10_}
\begin{aligned} \left\langle a \right\rangle = & \frac{\alpha \exp \left(-|\alpha|^{2} -\frac{\kappa}{2} t - i (\Delta - \frac{g_{2}}{2}) t \right)}{N(t)} \\ & \times \sum_{m=0}^{\infty} \left| p_{m} \right|^{2} \exp \left( -m \gamma t + i t \left( \Lambda_{1} + (g_{2} + \Lambda_{2}) m \right) \right) \\ & \times \exp \left( i t \Lambda_{3} m^{2} + |\alpha|^{2} e^{- t \kappa + 2 i (\Lambda_{1} + \Lambda_{2} m + \Lambda_{3} m^{2}) t} \right)
\end{aligned}
\end{equation}

\begin{figure}[ht] \centering \includegraphics[width=6 cm]{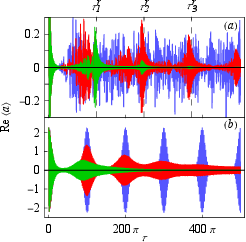} \caption{Influence of nonlinearities on the real part of the amplitude  $ \left\langle a\right\rangle  $  of the optical field at different relaxation rates. All parameters are normalized with respect to $ \Omega  $:  $ \tau =\Omega \, t $,  $ \gamma =0.01\kappa  $, $ \Delta =1 $,  $ g_{1} =0.1 $,  $ T=10.0{\rm \; mK} $, and  $ |\alpha |^{2} =5 $. Blue, red and green lines show the results obtained at  $ \kappa =\gamma =0 $, $ \kappa =0.0033 $  and  $ \kappa =0.01 $, respectively. (a)  $ g_{2} =0.0075 $,  $ g_{3} =0.00083 $. (b)  $ g_{2} =g_{3} =0 $.} \label{fig:Fig2} \end{figure}

Fig. \ref{fig:Fig2} demonstrates the influence of nonlinearities on the real part of the amplitude of the optical field at different relaxation rates. Since the used equations are considered in a coordinate system rotating at the driving field frequency, they do not include the cavity and driving field frequencies themselves, but their detuning, which is comparable in magnitude to the oscillation frequency of the mechanical resonator. Therefore, all parameters in the calculations are normalized to this frequency and are dimensionless. The cases when  $ g_{1} \ne g_{2} \ne g_{3} \ne 0 $  and  $ g_{1} \ne 0 $,  $ g_{2} =g_{3} =0 $  are presented in Figs. \ref{fig:Fig2}\emph{a} and \ref{fig:Fig2}\emph{b}, respectively. In the presence of only the Kerr interaction (the first term in Eq. \eqref{GrindEQ2}) for  $V_{eff}^{(2)} $), the amplitude of revivals increases with a decrease in the relaxation rates and their structure becomes clearer (Fig. \ref{fig:Fig2}\emph{b}). Fig. \ref{fig:Fig2}\emph{a} shows that the cross-Kerr interaction of the fourth degree (the second term in  $V_{eff}^{(1)} $) and two cross-Kerr interactions of the sixth and eighth degrees in the amplitudes (the second and third terms in  $V_{eff}^{(2)}$) led to the non-trivial dependence of  $Re\left\langle a\right\rangle$  on the relaxation rates. Relaxation processes result in the suppression of high-order harmonics.

The time of occurrence of revivals caused by each of the cross-Kerr terms depends on the values of the corresponding constants,  $g_{2}$, $\Lambda _{1}$, $\Lambda _{2}$, and $\Lambda _{3}$. Revivals, the origin of which is associated with the sixth- and eighth-degree cross-Kerr terms in amplitudes ($ \Lambda _{2}$ and $\Lambda _{3}$), appear later than revivals caused by the action of the fourth-degree Kerr and cross-Kerr terms in amplitudes ($g_{2}$ and $\Lambda _{1}$). If the largest of the relaxation times is shorter than the times of occurrence of revivals, the corresponding revivals will be strongly suppressed by the relaxation. The contribution of the  $ n $ - and  $ m $ -fold harmonics of the Kerr and cross-Kerr constants (see Eqs. \eqref{GrindEQ__9_}) and \eqref{GrindEQ__10_}) to the temporal behavior of the averaged amplitude of the optical mode is rather quickly leveled out by the rapidly decaying relaxation factors  $e^{-\kappa t}$  and  $e^{-\gamma t}$. Moreover, the intensity of harmonics from the sixth- and eighth-degree cross-Kerr terms is significantly smaller, than from the fourth-degree Kerr and cross-Kerr terms. It is therefore not surprising that at very slow relaxation the temporal behavior of the averaged optical amplitude becomes complex. The regular revival structure is largely destroyed due to the contribution of many harmonics from the sixth- and eighth-degree cross-relaxation terms. This behavior of the optomechanical system under consideration with an increase in its statistical properties with the slowing down of relaxation will be illustrated below using Poincar\'{e} sections.

Note that the effect of collapse and revival of the optical field in the optomechanical system is realized similarly to the collapse and revival of Rabi oscillations in the Jaynes-Cummings model for in cavity quantum electrodynamics \cite{pp35, pp36}.

\begin{figure}[ht]
    \centering
    \includegraphics[width=6 cm]{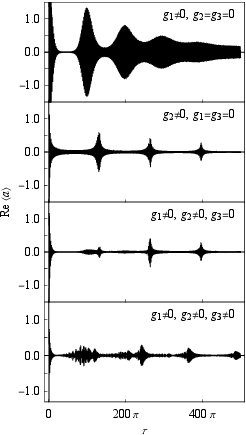}
    \caption{Effect of the nonlinearities of different orders on the temporal behavior of the real part of the amplitude $\left\langle a\right\rangle $. The normalized parameters are: $\tau =\Omega \, t$, $\kappa =0.0033$, $\gamma =0.01\kappa $, $\Delta =-1$, $g_{1} =0.1$, $g_{2} =0.0075$, $g_{3} =0.00083$, $T=10.0\, mK$, and $|\alpha |^{2} =5$.}
    \label{fig:FigNew1}
\end{figure}

Fig. \ref{fig:FigNew1} demonstrates how successive addition of nonlinear interactions modifies the collapse-revival structure. If only a linear interaction $g_{1} $ is present, the first collapse and revival are clearly manifested, while subsequent collapses and revivals are blurred. In the presence of only quadratic interaction $g_{2} $, a sequence of clearly localized revivals is formed. In the case of linear and quadratic interactions ($g_{1} \ne 0$, $g_{2} \ne 0$, $g_{3} =0$), the times of appearance of revivals almost coincide with those in the case of only quadratic interaction, but a tendency towards the appearance of low-intensity satellites is already emerging. The structure of revivals changes dramatically when all three optomechanical interactions ($g_{1} \ne 0$, $g_{2} \ne 0$, $g_3\neq 0$) come into play: the time of revival appearance is reduced, and each revival is blurred and acquires well-defined satellites.

\begin{figure}[ht]
    \centering
    \includegraphics[width=6 cm]{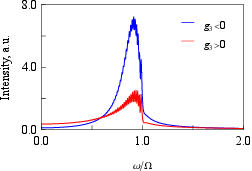}
    \caption{ The intensity spectrum for positive and negative values of the coupling strength $g_3$ of the model system. The normalized parameters are: $\tau =\Omega \, t$, $\kappa =0.02$, $\gamma =0.01\kappa $, $\Delta =1$, $g_{1} =0.1$, $g_{2} =0.0075$, $T= 10.0 mK$, and $| \alpha |^{2} =5$.  $g_{3} =0.00083$ (the red line) and $g_{3} =-0.00083$ (the blue line).}
    \label{fig:FigNew2}
\end{figure}

Fig. \ref{fig:FigNew2} shows the effect of the cubic interaction sign on the intensity (square of amplitude) spectrum  of the optical field in the cavity with all other parameters being equal. As can be seen, the  intensity increases more than 3 times when positive $g_{3} $ is replaced by $-g_{3} $.

\begin{figure}[ht] \centering \includegraphics[width=6 cm]{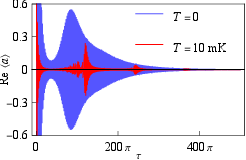} \caption{Effect of temperature on the temporal behavior of the real part of the amplitude  $ \left\langle a\right\rangle  $  of the model system. The normalized parameters are: $ \tau =\Omega \, t $   $ \kappa =0.01 $,  $ \gamma =0.01\kappa  $,  $ \Delta =1 $,  $ g_{1} =0.1 $,  $ g_{2} =0.0.0075 $,  $ g_{3} =0.00083 $, and  $ |\alpha |^{2} =5 $.} \label{fig:Fig3} \end{figure}

The temperature dependence of the temporal behavior of the optical amplitude is shown in Fig. \ref{fig:Fig3}. The presented behavior demonstrates that multiple revivals disappear at zero temperature as it follows from Eq. \eqref{GrindEQ__10_}.

\begin{figure}[ht] \centering \includegraphics[width=6 cm]{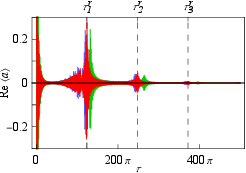} \caption{Effect of the cross-Kerr interaction with the parameter  $ \Lambda _{3}  $  on the temporal behavior of the real part of the amplitude  $ \left\langle a\right\rangle  $. The normalized parameters are:  $ \tau =\Omega \, t $   $ \kappa =0.01 $,  $ \gamma =0.01\kappa  $,  $ \Delta =1 $,  $ g_{1} =0.1 $,  $ g_{2} =0.0075 $,  $ T=10.0\, mK $, and  $ |\alpha |^{2} =5 $. The blue and green lines show the results obtained at  $ g_{3} =0.00083 $  and  $ g_{3} =0.0 $, respectively. The red line presents the results obtained at  $ g_{3} =0.00083 $  without taking into account the influence of the Kerr cross-coupling constant  $ \Lambda _{3}  $.} \label{fig:Fig4} \end{figure}

It can be seen from Fig. \ref{fig:Fig4} that for small values  $g_{3}$  the parameter  $ \Lambda _{3} \sim g_{3}^{2}  $  does not have a significant effect on the temporal behavior of the amplitude  $ \left\langle a\right\rangle  $. As  $ g_{3}  $  increases, the influence of this parameter becomes more noticeable and, as a result, the structure of revivals is changed.

Let us carry out an approximate analytical evaluation of the expression for  $ \left\langle a\right\rangle  $. For sufficiently large  $ t $  such that  $ e^{-\kappa t} \ll 1 $, the last exponential factor under the sum in Eq. \eqref{GrindEQ__10_} can be expanded in a Taylor series in  $ e^{-\kappa t}  $  and, restricting ourselves to the first two terms of the expansion, we obtain

\begin{equation} \label{GrindEQ__11_}
\begin{alignedat}{2}
\exp \left( |\alpha|^{2} e^{-\kappa t} e^{2i(\Lambda_{1} + \Lambda_{2} m + \Lambda_{3} m^{2}) t} \right) & \approx 1 + |\alpha|^{2} e^{-\kappa t} \\
\times e^{2i(\Lambda_{1} + \Lambda_{2} m + \Lambda_{3} m^{2}) t}.
\end{alignedat}
\end{equation}

Since  $ \Lambda _{3}\ll\Lambda _{2}\ll\Lambda _{1}  $, in the zeroth approximation the parameter  $ \Lambda _{3}  $  can be neglected. Note that the last approximation, however, does not exclude the influence of the cubic nonlinearity  $ g_{3}  $  of the optomechanical resonator, since  $ g_{3}  $  is also included in the definition of the Kerr  $ \Lambda _{1}  $  and cross-Kerr  $ \Lambda _{2}  $  parameters. Taking into account the indicated approximations, Eq. \eqref{GrindEQ__10_} can be written as

\begin{equation} \label{GrindEQ__12_} \begin{split} \left\langle a\right\rangle \approx & \frac{\alpha }{N(t)} \exp \left(-|\alpha|^{2} + (\gamma - \frac{\kappa }{2} )t - i \left(\Delta - \frac{g_{2} }{2} - \Lambda_{1} \right) t \right) \\ & \times \left[ \frac{1}{e^{\gamma t} \left( 1 + \bar{m} \right) - e^{i \left(g_{2} + \Lambda_{2} \right) t} \bar{m}} \right. \\ & \left. + \frac{|\alpha|^{2} e^{-\kappa t + 2i \Lambda_{1} t}}{e^{\gamma t} \left( 1 + \bar{m} \right) - e^{i \left(g_{2} + 3 \Lambda_{2} \right) t} \bar{m}} \right]. \end{split} \end{equation}

The terms in square brackets are responsible for the formation of the collapses and revivals. The second term quickly decays and affects the signal only at sufficiently short times ($ t \lesssim 1/\kappa  $). Analysis of the expression in square brackets of Eq. \eqref{GrindEQ__12_} allows us to determine the formation times of revivals (the first term). The second term tends to zero with increasing the relaxation rate of the optical mode, which leads to suppression of high-order harmonics. Consequently, the time moments of revivals in the optical field amplitude are approximately described as
\begin{equation} \label{GrindEQ__13_}
t_{k} \approx \frac{2k\pi }{g_{2} +\Lambda _{2} } ,
\end{equation}
where  $ k $  is the revival number. The collapse time  $ t_{c}  $  can approximately be found from the following formula:
\[\frac{\exp (-2|\alpha |^{2} \sin ^{2} (\Lambda _{1} t_{c} ))}{\sqrt{1+2\bar{m}\left(1+\bar{m}\right)\left(1-\cos \left(g_{2} t_{c} \right)\right)} } =\frac{1}{2} .\]
When deriving this formula, in Eq. \eqref{GrindEQ__10_} terms  $ g_{2}^{2}$ and $g_{3}^{2}  $  of a higher order of smallness were neglected, but  $g_{1},\,g_{1}^{2}$, and $g_{2}$  were taken into account.

Now let us compare the calculations using the approximate equation \eqref{GrindEQ__12_} and the original equation \eqref{GrindEQ__10_}, which we will henceforth call ``exact''. Fig. 5 shows that these calculations agree well with each other.

\begin{figure}[htp] \centering \includegraphics[width=6 cm]{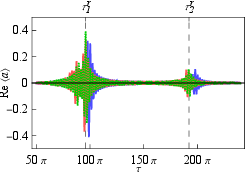} \caption{Comparison of calculations using the approximate Eq. \eqref{GrindEQ__12_} and the ``exact'' Eq. \eqref{GrindEQ__10_}. The normalized parameters are:  $ \tau =\Omega \, t $,  $ \tau _{k}^{r} =\Omega t_{k}  $,  $ \kappa =0.01 $,  $ \gamma =0.01\kappa  $,  $ \Delta =1 $,  $ g_{1} =0.1 $,  $ g_{2} =0.01$,  $ T=10.0{\rm \; mK} $, and  $ |\alpha |^{2} =5 $. The red and blue lines show the ``exact'' results obtained at  $ g_{3} =0.00083$  and  $ g_{3} =0.0 $, respectively. The green line presents the approximate results obtained at  $ g_{3} =0.00083 $  without taking into account the influence of the Kerr cross-coupling constant  $\Lambda _{3}$.} \label{fig:Fig5}\end{figure}

\begin{figure}[htp] \centering \includegraphics[width=6 cm]{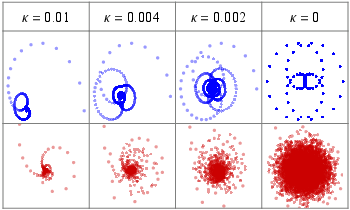} \caption{Poincar\'{e} sections for different relaxation rates. All parameters are normalized with respect to $ \Omega  $: $ \gamma =0.01\kappa  $,  $ T=10.0\; mK $,  $ |\alpha |^{2} =5 $,  $ g_{1} =0.1 $,  $ \Delta =2 $. The top row (blue points) presents the results obtained for the optomechanical system only with the Kerr nonlinearity ($ g_{2} =g_{3} =0 $). The botton row (red points) shows the results obtained for the optomechanical system with the Kerr nonlinearity taking into account the cross-Kerr nonlinearities ($ g_{2} =0.0075 $,  $ g_{3} =0.00083 $).} \label{fig:Fig6} \end{figure}

We will analyze the dynamics of the system under consideration using Poincar\'{e} sections, which are a set of points in phase space and allow us to study the behavior of our system in more detail \cite{pp37}. Fig. \ref{fig:Fig6} shows a series of Poincar\'{e} sections where  $ \left\langle a\right\rangle  $  and  $ \left\langle da/dt\right\rangle  $  are chosen as coordinates in the phase space. The scale is chosen so that the structure of the sections is better visible. The top row of Poincar\'{e} sections corresponds to the behavior of the optomechanical system only with the Kerr nonlinearity, and the bottom row shows peculiarities in the behavior due to the cross-Kerr nonlinearities. It is evident that the cross-Kerr nonlinearities lead to the fact that with a decrease in relaxation rates the structure of the Poincar\'{e} sections is disrupted.
As we noted above, with a decrease in the relaxation rates of the optical and mechanical subsystems, the contribution of the  $ n $ - and  $ m $ -fold harmonics of the Kerr and cross-Kerr constants becomes increasingly evident in the dissipative dynamics of the averaged amplitude of the optical mode. As a result, the regular revival structure is destroyed and replaced by a disordered multi-frequency one. Thus, the optomechanical system increasingly moves from dynamic behavior to statistical behavior, which is illustrated by the Poincar\'{e} sections in the bottom row in Fig. \ref{fig:Fig6}. The top row corresponds to the situation when the dissipative dynamics of the system under study is due to the Kerr interaction $ \Lambda _{1} (g_{1} ) $. In this case, the revival structure is preserved and is not radically transformed, when the relaxation rate  $ \kappa  $  changes.

The predicted new properties of collapses and revivals, such as the reduction in the time of appearance of revivals, their blurring and the appearance of their well-defined satellites, can be tested experimentally in optomechanical systems with all three of the above-considered couplings. Such couplings can be realized in optomechanical systems, which use optics with membrane-in-the-middle configurations \cite{pp38} and superfluid optomechanics \cite{pp39} as well as in an optomechanical system consisting of a drumhead nano-electro-mechanical resonator coupled to a microwave cavity \cite{pp31}.

\section{Fano interference}

\begin{figure}[ht]
    \centering
    \includegraphics[width=8 cm]{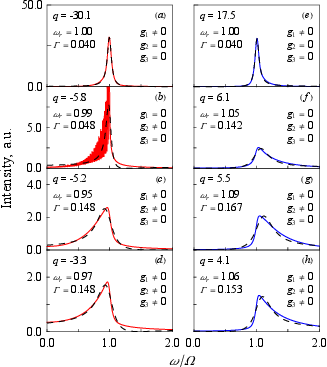}
    \caption{Effect of the nonlinearities of different orders on the intensity spectrum. The normalized parameters are: $\tau =\Omega \, t$, $\kappa =0.08$, $\gamma =0.0002$, $\Delta =1$ (the left column), $\Delta =-1$ (the right column), $g_{1} =0.1$, $g_{2} =0.0075$, $g_{3} =0.00083$, $T=10.0\, mK$, and $|\alpha|^{2} =5$. The dashed lines represent the result of fitting using formula \eqref{GrindEQ__14_}.}
    \label{fig:FigNew3}
\end{figure}

Now we will consider the spectral dependence of the intensity  of the optical field, $I(\omega )=\left|\langle a(\omega )\rangle \right|^{2} $. It turned out (Fig. \ref{fig:FigNew3} shows) that this dependence has the form (at least with a certain choice of parameters) of an asymmetric spectral line, similar to the Fano resonance line-shape \cite{ppFano}. This indicates that the dissipative dynamics of the optomechanical system can be related to the physics of Fano resonances. Indeed, Fano resonance is a fairly common phenomenon observed for various physical systems (see, for example, \cite{ppLimonov, ppLebedev}). According to the original source \cite{ppFano}, Fano resonance is realized in a system in which a discrete quantum state interferes with a continuum of other states of the medium. The asymmetric Fano resonance line can be written as

\begin{equation} \label{GrindEQ__14_}
I(\omega )\sim \frac{(q\Gamma +\omega -\omega _{r} )^{2} }{(\omega -\omega _{r} )^{2} +\Gamma ^{2} },
\end{equation}

\noindent where $q$ is the Fano parameter, characterizing the asymmetry of the line-shape, $\omega _{r} $ and $\Gamma $ are the spectral position and line width, respectively. If $q\gg 1$, the Fano profile takes the form of a symmetric Lorentzian line.

It was shown that a quantum mechanical system having a discrete level and interacting with an energy continuum of other states can be in many ways analogous to the forced oscillation of coupled classical oscillators, for which a similar Fano spectral line-shape is observed \cite{ppJoe, ppRiffe, ppIizawa, ppLebedev}.  It is obvious that in our case the formation of the Fano line-shape is caused by the interference effect of two quantum nonlinearly interacting damped modes -- optical cavity mode and mechanical resonator mode. Due to the difficulty of solving such a quantum nonlinear problem, we can give here only an intuitive explanation of the physical nature of the Fano resonance, assuming that it is formed due to the destructive interference of two energy transfer channels \cite{ppJoe, ppRiffe, ppIizawa, ppLebedev} from a pulsed laser source to an optomechanical system coupled by cross-Kerr interactions (in a coordinate system rotating with the frequency $\omega _{d} $ of the pulsed source) through the frequencies of the optical and mechanical modes. Fig. \ref{fig:FigNew3} illustrates that if $\Delta =\omega _{c} -\omega _{d} >0$ (``red'' region of the spectrum), destructive interference occurs on the high-frequency wing of the spectral line, and when $\Delta =\omega _{c} -\omega _{d} <0$ (``blue'' region of the spectrum) such interference occurs on the low-frequency wing.

The series of graphs in Fig. \ref{fig:FigNew3} shows the spectral dependence of the intensity $\left|\langle a(\omega )\rangle \right|^{2} $ of the optical field for optomechanical interactions of different orders in mechanical displacements: linear (\emph{a}, \emph{e}), quadratic (\emph{b}, \emph{f}), linear and quadratic (\emph{c}, \emph{g}), as well as linear, quadratic and cubic (\emph{d}, \emph{h}). The graphs of the left and right columns were obtained for the case of ``red'' ($\Delta =1$) and ``blue'' ($\Delta =-1$) tuning, respectively. Spectral dependences of intensities (solid lines) were calculated using the Fourier transforms of equation \eqref{GrindEQ__9_} for the selected parameters and were fitted with Fano profiles (dashed lines). The shape of the spectral lines for $\Delta =1$ (\emph{b}, \emph{c}, \emph{d}) is asymmetric, i.e. the decrease in intensity on the high-frequency wing  occurs faster than on the low-frequency one. In this case, the Fano parameter $q$ is negative, and the small value of its modulus indicates a relatively large asymmetry of the spectral line. For the spectral line in graph (\emph{a}), the value of the parameter $\left|q\right|\gg 1$ and the line shape is almost symmetrical, close to Lorentzian. The values of resonant frequencies and spectral line widths are given in each plot. For the case $\Delta =-1$, the spectral lines  (\emph{f}, \emph{g}, \emph{h}) are also asymmetrical, but their profiles on the low-frequency wing are steeper, the Fano parameter $q$ is positive and small in value, characteristic of strongly asymmetrical lines.For the red-detuned optical field, the normalized Fano resonance frequency is 1–5\% less than 1, and for the blue-detuned field, it is 5–9\% greater than 1. The spectral line for linear optomechanical interaction (\emph{a}, \emph{e}) is fitted by the Fano profile with a positive parameter $q\gg 1$ and, accordingly, has a practically symmetric, Lorentzian shape.

Thus, the presence of asymmetry of the spectral line of the optical field intensity in the cavity, being a characteristic feature of Fano interference, is most clearly manifested with an increase in the degree of nonlinearity. As can be seen from Fig. 9, the Fano parameter $q$ decreases from 17.5 to 4.1 (at $\Delta =-1$) and in modulus from 30.1 to 3.3 (at $\Delta =1$).

The asymmetric shape of Fano resonances allows them to be easily identified as noticeable features in experimental spectra. On the other hand, the fact of the presence of asymmetry and a steep decline/rise of the spectral line provides additional opportunities for the development of high-precision meso- or nanoscale sensors using the models of optomechanical systems considered here.

\section{Conclusions}

We have studied  optomechanical systems, in which, in addition to the linear interaction, quadratic and cubic interactions with the mechanical resonator are taken account. In the framework of the nonsecular perturbation theory, using the Bogoliubov averaging method, we have constructed the effective Hamiltonian of these systems in which the Kerr self-action and three cross-Kerr interactions arise with the fourth, sixth and eighth degrees in the amplitudes of the optical field and mechanical vibrations, respectively. The intensities of the Kerr and cross-Kerr terms in the effective Hamiltonian depend on all three constants of initial optomechanical interactions. These nonlinearities form the energy spectrum of the optomechanical system which creates the possibility of realizing the photon blockade. We have considered the dynamical behavior of the optomechanical system when at the initial time the optical mode is populated in a monochromatic coherent state  $ {\left| \alpha  \right\rangle}  $, and the mechanical resonator is in some thermal state. The possibility of collapses and revivals in the mean amplitude of the optical field was shown. The main contribution to the formation of the regular structure in the revival sequence is provided by the Kerr self-action of the optical mode and the cross-Kerr interaction of the fourth degree in optical and mechanical amplitudes. The cross-Kerr interactions of the sixth- and eighth-order degrees in amplitudes destroy the regular structure of revivals. With an increase in the decay rate of the optical mode, high-frequency harmonics in the averaged amplitude of the optical field are suppressed and the sixth- and eighth-order cross-Kerr nonlinearities cease to have an impact the behavior of the dynamic system. The influence of these cross-Kerr nonlinearities is also completely suppressed at zero temperature. The dynamics of the system under consideration was additionally analyzed using Poincar\'{e} sections. Poincare sections and the transition of a nonlinear optomechanical system to quasi-periodic or chaotic dynamics were studied in the classical approximation \cite{ppSaiko2025}.

It is shown that the dissipative dynamics of the optomechanical system can be related to the physics of Fano resonances. The spectral line of the optical field
intensity in the cavity has a pronounced asymmetric shape, which is a characteristic feature of Fano interference. The asymmetry of the spectral line is
most pronounced with an increase in the degree of nonlinearity.

At present, the few results of experimental studies \cite{pp29} of the limit cycle dynamics of microwave optomechanical system in the self-oscillation regime within the classical approximation have shown the importance of taking into account nonlinear interactions up to the third order in mechanical displacements. Possible examples of other optomechanical systems with nonlinear couplings, which use optics with membrane-in-the-middle configurations and superfluid optomechanics, are also discussed in \cite{pp29}. Our results concern purely quantum aspects of the behavior of such nonlinear optomechanical systems and can be used to create extremely sensitive detectors as well as devices with controlled nonlinear couplings between individual photons and phonons for potential applications in quantum information processing.

\end{document}